\documentstyle[prd,aps,floats,tabularx,epsfig]{revtex}
\newcommand{\be}{\begin{equation}}
\newcommand{\ee}{\end{equation}}
\newcommand{\bea}{\begin{eqnarray}}
\newcommand{\eea}{\end{eqnarray}}

\begin{document}
\setlength{\unitlength}{1mm}
\twocolumn[\hsize\textwidth\columnwidth\hsize\csname@twocolumnfalse\endcsname
\title{Current constraints on the dark energy equation of state}
\author{Rachel Bean$^\sharp$ and Alessandro Melchiorri$^\flat$}
\address{ $^\sharp$ Theoretical Physics, The Blackett Laboratory, Imperial
 College, Prince Consort Road, London, U.K.\\
$^\flat$ NAPL, University of Oxford, Keble road, OX1 3RH, Oxford, UK}
\maketitle
\begin{abstract}
We combine complementary datasets from Cosmic Microwave Background (CMB) anisotropy measurements, high redshift supernovae (SN-Ia) observations and data from
local cluster abundances and galaxy clustering (LSS) to 
constrain the dark energy equation of state
parameterized by a constant pressure-to-density ratio $w_Q$.
Under the assumption of flatness, we find 
$w_Q < -0.85$ at $68 \%$ c.l., providing no significant evidence for quintessential behaviour different from that of a cosmological constant.
We then generalise our result to show that the constraints placed on a constant $w_{Q}$ can be safely extended to dynamical theories. We consider a variety of quintessential dynamical models based on 
inverse power law, exponential and oscillatory scaling potentials. We find that SN1a observations are `numbed' to dynamical shifts in the equation of state, making the prospect of reconstructing $w(z)$, a challenging one indeed.
\end{abstract}
\bigskip]

{\it Introduction.}
The discovery that the universe's evolution may be dominated by an effective cosmological constant  \cite{super1}, is one of the most remarkable cosmological findings of recent years. An exceptional opportunity is now opening up to decipher the nature of dark matter \cite{dark}, to test the veracity of theories and reconstruct the dark matter's profile using a wide variety of observations over a broad redshift range.

One matter candidate that could possibly explain the observations is a dynamical scalar ``quintessence'' field. One of the strong aspects of quintessence theories is that they go some way to explaining the fine-tuning problem, why the energy density producing the acceleration is $\sim 10^{-120}M_{pl}^{4}$. A vast range of ``tracker'' (see for example \cite{quint,brax}) and ``scaling'' (for example \cite{wett}-\cite{ferjoy}) quintessence models exist which approach attractor solutions, giving  the required energy density, independent of initial conditions. The common characteristic of quintessence models is that their equations of state,$w_{Q}=p/\rho$, vary with time whilst a cosmological constant remains fixed at $w_{Q=\Lambda}=-1$. Observationally distinguishing a time variation in the equation of state or finding $w_Q$ different from $-1$ will therefore
be a success for the quintessential scenario.

In this paper we will combine the latest observations of the 
Cosmic Microwave Background (CMB) anisotropies provided by
the Boomerang \cite{Boom2}, DASI \cite{Dasi} and Maxima \cite{Maxima} 
experiments and the information from Large
Scale Structure (LSS) with the luminosity distance of high 
redshift supernovae (SN-Ia) to put constraints on the dark energy 
equation of state parameterised by a redshift independent quintessence-field pressure-to-density ratio $w_Q$.We will also make use of the Hubble Space Telescope (HST) constraint 
on the Hubble parameter $h=0.72 \pm 0.08$ \cite{freedman}. We will then also consider whether one can feasibly extract information about the time variation of $w$ from observations.


The importance of combining different data sets in order to obtain 
reliable constraints on $w_Q$ has been stressed by
many authors (see e.g. \cite{PTW}, \cite{hugen},\cite{jochen}), 
since each dataset suffers from degeneracies between the various
cosmological parameters and $w_Q$ . Even if one restricts consideration
 to flat universes and to a value of $w_Q$ constant
with time the SN-Ia luminosity distance and position of the first CMB peak are highly degenerate in $w_Q$ and $\Omega_Q$, the energy density in quintessence.

The paper is therefore structured as follows: in section II and III
we will present the CMB, SN-Ia and LSS data used in the analysis.
In section IV we will present the results of our analysis.
We will consider the implications for a dynamical $w_{Q}$
in section V. Section VI will be devoted to the
discussion of the result and the conclusions. 

\medskip
{\it Constraints from CMB.}
The effects of quintessence on the angular power spectrum of the
CMB anisotropies are several \cite{hugen,doran,AS}. In the class of models we are considering, however, with
a negligible value of $\Omega_Q$ in the early universe
  
in order to satisfy the BBN bound (\cite{bhm}), the effects
can be reduced to just two . 

Firstly, since the inclusion of quintessence
changes the overall content of matter and energy, the angular
diameter distance of the acoustic horizon size at recombination will be altered.
 
In flat models (i.e. where the energy density in matter is
equal to $\Omega_M=1-\Omega_Q$), 
this creates a shift in the
peaks positions of the angular spectrum as 
\begin{eqnarray}{\cal R}&=&\sqrt{(1-\Omega_Q)}y, \label{Req} \\
y&=&\int_0^{z_{dec}}
{[(1-\Omega_Q)(1+z)^3+\Omega_{Q}(1+z)^{3(1+w_Q)}]^{-1/2} dz}
\nonumber \end{eqnarray}
It is important to note that the effect is completely degenerate
in the interplay between $w_Q$ and $\Omega_Q$.
Furthermore, it does not qualitatively add any new
features additional to those produced by the presence of a
cosmological constant \cite{eb} and it is not highly sensitive
to further time dependencies of $w_Q$.

\begin{figure}[thb]
\begin{center}
\epsfxsize=7.2cm
\epsfysize=6.2cm
\epsffile{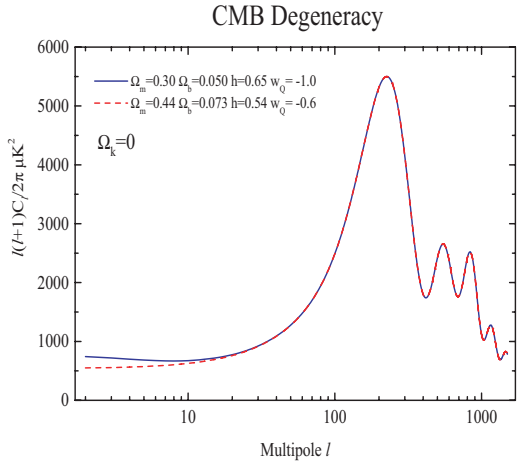}
\end{center}
\caption{CMB power spectra and the angular diameter distance 
degeneracy. The models are computed assuming flatness, 
$\Omega_k=1-\Omega_M-\Omega_Q=0$). The Integrated Sachs Wolfe effect on large angular scale slightly breaks the degeneracy. The degeneracy can be broken with a strong prior on $h$, in this paper we use the results from the HST.}
\label{figomega}
\end{figure}

Secondly, the time-varying Newtonian potential after decoupling will
produce anisotropies at large angular scales through the Integrated
Sachs-Wolfe (ISW) effect. The curve in the CMB angular spectrum
on large angular scales depends not only on the value of
$w_Q$ but also its variation with redshift.
However, this effect will be difficult to
disentangle from the same effect generated by a cosmological constant,
especially in view of the affect of cosmic variance and/or
gravity waves on the large scale anisotropies.

In order to emphasize the importance of degeneracies between
all these parameters while analyzing the CMB data, we plot
in Figure 1 some degenerate spectra, obtained keeping
the physical density in matter $\Omega_Mh^2$, the physical density
in baryons $\Omega_bh^2$ and
${\cal R}$ fixed. As we can see from the plot, models degenerate
in $w_Q$ can be constructed.
However, as we will utilise in the next sections, the combination
of the different datasets can break the mentioned degeneracies.

To constrain $w_Q$ from CMB, we
perform a likelihood analysis comparing the recent
CMB observations with a set of models with cosmological parameters
sampled as follows: $0.1 < \Omega_{m} < 1.0$, 
$-1.0 \le w_Q \le -0.55$,
$0.015 < \Omega_{b} < 0.20$;
$0< \Omega_{Q} < 0.9$ and $0.45 < h < 0.95$.  We vary the
spectral index of the primordial density perturbations within the
range $n_s=0.60, ..., 1.40$, we allow for a possible
reionization of the intergalactic medium by varying
the CMB photons optical depth in the range
$0.0 < \tau_C < 0.4$  and we re-scale the fluctuation amplitude
by a pre-factor $C_{10}$, in units of $C_{10}^{COBE}$.  We also
restrict our analysis to {\it flat} models, $\Omega_{tot}=1$, and we
add a conservative  external prior on the age of the 
universe $t_0 > 10$ Gyrs (see e.g. \cite{FMS}).

\begin{figure}[thb]
\begin{center}
\epsfxsize=6.2cm
\epsfysize=6.2cm
\epsffile{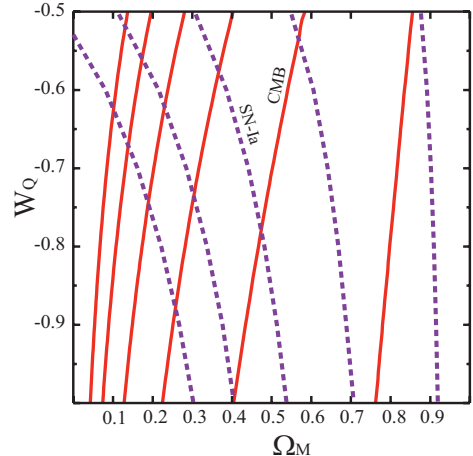}
\end{center}
\caption{Contours of constant $R$ (CMB) and 
$SN-Ia$ luminosity distance in the $w_Q$-$\Omega_M$ plane.
The degeneracy between the two distance measures can be broken 
 by combining the two sets of complementary information. 
The luminosity distance is chosen to be equal to $d_{l}$ at
$z=1$ for a fiducial model with 
$\Omega_{\Lambda}=0.7$, $\Omega_{M}=0.3$,$h=0.65$.
(We note that as $\Omega_{Q}=1-\Omega_M$ 
goes to zero the dependence of $R$ and $d_{L}$ upon 
$w_{Q}$ also become zero, as there is no dark energy present.)}
\label{figomega2}
\end{figure}
\medskip

In order to speed-up the computation time of the theoretical models
for different $w_Q$ we make use of a $k$-splitting 
technique \cite{teggy}.
Basically a $w_Q=-1$ and $w_Q >-1$ model are calculated in two
different ways. For low $\ell$ the 2 models are computed
in the ordinary way by solving the Boltzmann equation, in order
to properly take in to account the ISW effect. 
For the high $\ell$ just a flat, $w_Q=-1$ model is
calculated.  This $w_Q=-1$ model is then shifted using the 
expression for the angular diameter distance in equation (1) 
to obtain the $w_Q >-1$ models.

The theoretical models are computed using a modified version of 
the publicly available {\sc
cmbfast} program (\cite{CMBFAST}) and are compared with the recent
BOOMERanG-98, DASI and MAXIMA-1 results.  The power spectra from these
experiments were estimated in $19$, $9$ and $13$ bins respectively,
spanning the range $25 \le \ell \le 1100$.  We approximate the
experimental signal $C_B^{ex}$ inside the bin to be a Gaussian
variable, and we compute the corresponding theoretical value
$C_B^{th}$ by convolving the spectra computed by CMBFAST with the
respective window functions. When the window functions are not
available, as in the case of Boomerang-98, we use top-hat window
functions.  The likelihood for a given cosmological model is then
defined by $-2{\rm ln}
{\cal L}=(C_B^{th}-C_B^{ex})M_{BB'}(C_{B'}^{th}-C_{B'}^{ex})$ where
$C_B^{th}$ ($C_B^{ex}$) is the theoretical (experimental) band power
and $M_{BB'}$ is the Gaussian curvature of the likelihood matrix at
the peak.  We consider $10 \%$, $4 \%$ and $4 \%$ Gaussian distributed
calibration errors (in $\mu K$)for the BOOMERanG-98, DASI and MAXIMA-1
experiments respectively and we take in to account for the beam error
in BOOMERanG-98 by analytic marginalization as in \cite{bridle}.
We also include the COBE data using Lloyd Knox's RADPack packages.

\medskip
{\it Constraints from Supernovae.}
Evidence that the universe's expansion rate was accelerating was first 
provided by two groups, the SCP and High-Z Search Team(\cite{super1}) using 
type Ia supernovae (SN-Ia) to probe the nearby expansion dynamics. 
SN-Ia are good standard candles, as they exhibit a strong phenomenological 
correlation between the decline rate and peak magnitude of the 
luminosity. The observed apparent bolometric luminosity 
is related to the luminosity distance, measured in Mpc, by
$m_{B}=M+5 log d_{L}(z)+25$.
where M is the absolute bolometric magnitude. 
The luminosity distance is sensitive to the cosmological 
evolution through an integral dependence on the Hubble factor 
$d_{l}=(1+z)\int_{0}^{z} (dz'/H(z',\Omega_{Q},w_{q})$ 
and therefore can be used to constrain the scalar equation of state. 
We evaluate the dark energy $\Omega/w$ likelihoods 
assuming a constant equation of state, such that 
$H(z)=\rho_{0}\sum_{i}\Omega_{i}(1+z)^{(3+3w_{i})}$. 
The predicted $m_{eff}$ is then calculated by calibration 
with low-z supernovae observations \cite{calan} where 
the Hubble relation $d_{l}\approx H_{0}cz$ is obeyed. 
We calculate the likelihood, ${\cal L}$, using the relation 
${\cal L}={\cal L}_{0}\exp(-\chi^{2}(\Omega,w_Q)/2)$ where ${\cal L}_{0}$ 
is an arbitrary normalisation and $\chi^{2}$ is evaluated 
using the observations of the SCP group, 
marginalising over $H_{0}$. As can be seen in Figure \ref{figomega2} 
there is an inherent degeneracy in the luminosity distance in the 
$\Omega_{M}/w_{Q}$ plane; one can see that little can be found out 
about the equation of state from luminosity distance data alone. However,
 the degeneracies of CMB and SN1a data complement 
one another so that together they offer a more powerful approach 
for constraining $w_Q$. 

\medskip
\begingroup\squeezetable
\begin{table}[bt]
\renewcommand*{\arraystretch}{1.5}
\caption{
  Constraints on $w_Q$ and $\Omega_M=1-\Omega_Q$ 
  using different priors and datasets.
  We always assume flatness and $t_0>10$~Gyr.
  The $1\sigma$ limits are found from the 16\% 
  and 84\% integrals of the marginalized likelihood. 
  The HST prior is $h=0.72 \pm0.08$ while for the BBN prior
  we use the conservative bound $\Omega_bh^2=0.020\pm0.005$.
}\label{tab}
\begin{tabular}{lrrrr}
\hline
\\
CMB+HST
&$w_Q<-0.62$
\\
&$0.15<\Omega_M<0.45$
\\
CMB+HST+BBN
&$-0.95<w_Q<-0.62$
\\
&$0.15<\Omega_M<0.42$
\\
CMB+HST+SN-Ia
&$-0.94<w_Q<-0.74$
\\
&$0.16<\Omega_M<0.34$
\\
CMB+HST+SN-Ia+LSS
&$w_Q<-0.85$
\\
&$0.28<\Omega_M<0.43$
\\
\hline
\\
\end{tabular}
\end{table}
\endgroup

\begin{figure}[thb]
\begin{center}
\epsfxsize=6.2cm
\epsfysize=7.2cm
\epsffile{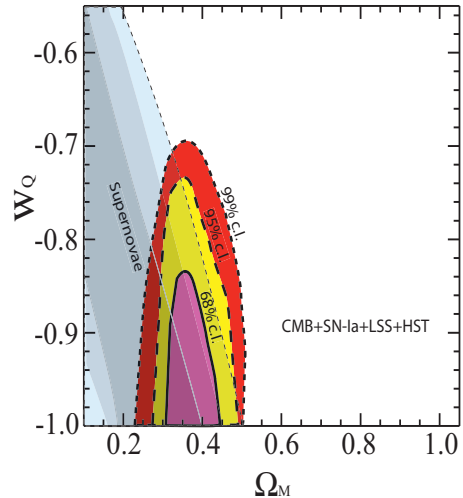}
\end{center}
\caption{The likelihood contours in the ($\Omega_M$, $w_Q$) plane,
with the remaining parameters taking their best-fitting values for the
joint CMB+SN-Ia+LSS analysis described in the text.
The contours correspond to 0.32, 0.05 and 0.01 of the peak value of the
likelihood, which are the 68\%, 95\% and 99\% confidence levels respectively.}
\label{figo}
\end{figure}
\medskip
{\it Results.}
Table I shows the $1$-$\sigma$ constraints on $w_Q$ for different 
combinations of priors, obtained after marginalizing
over all remaining {\it nuisance} parameters.
The analysis is restricted to {\it flat} universes.
One can see that $w_Q$ is poorly constrained from CMB data alone,
even when the HST strong prior on the Hubble parameter, 
$h=0.72\pm0.08$, is assumed.
Adding a Big Bang Nucleosynthesis prior, 
$\Omega_bh^2 =0.020 \pm 0.005$, has small effect on the CMB+HST 
result.
Adding SN-Ia breaks the CMB $\Omega_Q-w_Q$ degeneracy and 
improves the upper limit on $w_Q$, with $w_Q <-0.74$.
Finally, including information  from local cluster abundances through 
$\sigma_8=(0.55\pm0.1)\Omega_M^{-0.5}$, where $\sigma_8$ is the {\it rms} 
mass fluctuation in spheres of $8 h^{-1}$ Mpc, further breaks
the quintessential-degeneracy, giving $w_Q <-0.85$ at $1$-$\sigma$.
Also reported in Table I, are the constraints on $\Omega_M$.
As we can see, the combined data suggests the presence of dark
energy with high significance, even in the case CMB+HST.
It is interesting to project our likelihood in the $\Omega_Q-w_Q$ plane. 
Proceeding as in \cite{melk2k}, we attribute a likelihood to a point
in the ($\Omega_M$, $w_Q$) plane by finding the remaining parameters that
maximise it. We then define our $68\%$, $95\%$
 and  $99\%$ contours to be where the likelihood falls to $0.32$, $0.05$ and
$0.01$ of its peak value, as would be the case for a
two dimensional multivariate Gaussian.
In Figure 3 we plot likelihood contours in the
($\Omega_M$, $w_Q$) plane for the joint analyses of
CMB+SN-Ia+HST+LSS data together with the contours from the SN-Ia dataset
only. As we can see, the combination of the data breaks the
luminosity distance degeneracy.

\begin{figure}[thb]
\begin{center}
\epsfxsize=6.2cm
\epsfysize=6.2cm
\epsffile{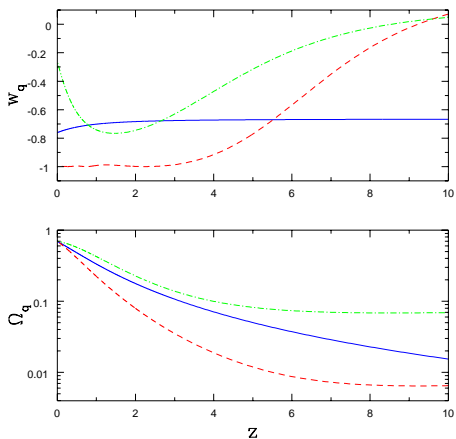}
\end{center}
\caption{The variation of $\Omega_{Q}$ and $w_{Q}$ with redshift for the three models described in section V. The power law potential (full line) $w_{Q}$ shows a steady small variation, the exponential feature potential, (short dashed line) acts remarkably like a cosmological constant at late times, after a deviation from behaving like normal matter with $w_{Q}\sim 0$, whilst the oscillating potential (dot-dash) shows a continual variation in $w_{Q}$.}
\label{rbfig1}
\end{figure}

\medskip
{\it Constraining dynamical models.}
In the previous section we obtained bounds on the equation
of state parameter $w_Q$ by assuming it is a constant,
independent of the redshift.
However, in quintessential models the equation of
state can vary with time.
It is therefore useful to discuss how well our constraints 
on $w_{Q}$ apply to less trivial models. 
There are a wide variety of quintessential models; 
we illustrate our analysis using representatives of three of the
most general classes of model:
the inverse power law,$V(\phi)=V_{0}/\phi^{p}$ \cite{quint} with p=1, 
an exponential scaling potential with a
feature, $V(\phi)=V_{0}e^{-\lambda\phi}(A+(\phi-\phi_{0})^{2})$ \cite{AS},
with $\lambda=10, A=0.008, \phi_{0}=25.8$,
and an oscillatory scaling
potential,$V(\phi)=V_{0}e^{-\lambda\phi}(1+A sin(\nu \phi))$ \cite{Dodelson:2000fq} with $\lambda=4,A=0.98,\nu=0.51$.
In each case $V_{0}$ is chosen so that $\Omega_{Q}=0.7$ and $H_{0}$ is 65. The particular time dependent characteristics of each are shown in Figure \ref{rbfig1}.

\begin{figure}[thb]
\begin{center}
\epsfxsize=6.2cm
\epsfysize=6.2cm
\epsffile{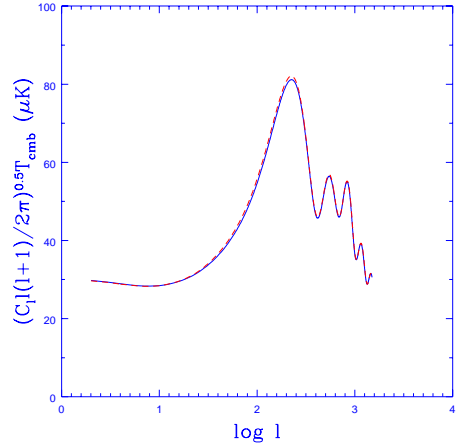}
\end{center}
\caption{Comparison of CMB temperature power spectra for the dynamical quintessence model with an exponential potential with a feature described in section V, and a model with constant $w_{Q}=w_{eff}$ for the dynamical model. In both cases with $\Omega_{Q}=0.7$ and $H_{0}=65$. One can see that the constant model is a remarkably good approximation to the dynamical model, despite the equation of state of the dynamical model varying significantly from the effective value from recombination until nowadays. }
\label{rbfig2}
\end{figure}

\begin{figure}[thb]
\begin{center}
\epsfxsize=6.2cm
\epsfysize=6.2cm
\epsffile{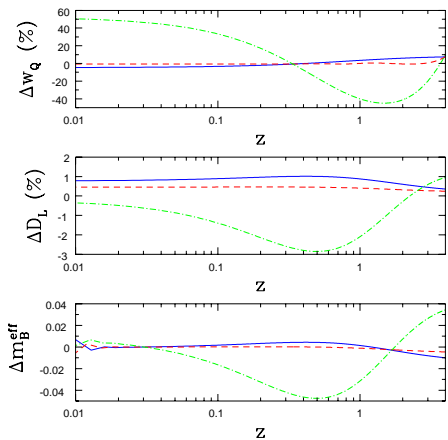}
\end{center}
\caption{The implication of a deviation $\Delta w_{Q}=1-w_{eff}/w_{Q}$ on $\Delta D_{L}=1-D_{L}(w_{eff})/D_{L}(w_{Q})$ and $\Delta m_{B}=m_{B}(w_{Q})-m_{B}(w_{eff})$ for the three models in section V.  The tracker potential (full line) has $w_{eff}=-0.727$, the exponential potential with feature (dashed line) has $w_{eff}=-0.993$ and the oscillatory potential (dot-dash) has $w_{eff}=-0.529$. Notice that $D_{L}$ and $m_{B}$ are in turn both desensitized to any variation in $w_{Q}$.}
\label{rbfig3}
\end{figure}

If we are to constrain dynamical models, 
we need to understand how well the effect of a time varying $w_{\phi}$ 
can be modeled by a constant $w_{eff}$. 
In \cite{huey} it was shown that in models in which the dark energy 
component is negligible at last scattering, the CMB and matter power 
spectra are well approximated by 
$w_{eff}=\int da \Omega_{Q}(a)w_{Q}(a)$. This is demonstrated in figure \ref{rbfig2} 
in the case of the exponential potential with a feature, using $w_{eff}=-0.993$. 
In \cite{bhm} we showed that if the dark energy component \emph{is} 
a significant proportion, as can be seen in scaling quintessence models, 
the dark energy component can be modelled as an additional contribution 
to the effective number of relativistic degrees of freedom. 
We restrict ourselves to former case in which $\Omega_{\phi}(MeV)$ is 
negligible. In \cite{PTW} it is noted that although $w_{eff}$ 
is a good measure for modelling CMB spectra it may not be such a 
good measure when considering $d_{l}$. We investigate 
whether this is actually the case using the three models outlined above, a power law, exponential 
scaling and scaling oscillatory potential, as test cases. 
In Figure \ref{rbfig3} we show that even a substantial 
deviation from $w_{eff}$ at late times produces a small change 
in $\rho_{Q}$. The effect on $m_{B}$ is doubly `numbed', 
firstly because of the smoothing by the integral relation with $H$ in $d_{l}$, see \cite{maor} for a previous discussion of this, and secondly because $m_{B}\sim log  d_{l}$. As a result the bolometric magnitude taken from the SN1a data is highly insensitive to variations in the equation of state. This doesn't bode well if we are to try and reconstruct the time varying equation of state from observations. It looks more likely that $w_{eff}$ will be a more tangible observable.     

\medskip
{\it Conclusions.}
In this paper we have provided new constraints on the dark energy 
equation of state parameter $w_Q$ by combining different 
cosmological data.
The new CMB results provided by Boomerang and DASI improve the constraints 
from previous and similar analysis (see e.g., \cite{PTW},\cite{bondq}), with
$w_Q<-0.85$ at $68 \%$ c.l. ($w_Q<-0.76$ at $95 \%$ c.l.).
We have also demonstrated how the combination of CMB data 
with other datasets is crucial in order to break 
the $\Omega_Q-w_Q$ degeneracy.
The constraints from each single datasets are, as expected, 
quite broad but compatible between each other, providing an important 
consistency test.
When comparison is possible (i.e. restricting to similar priors and
datasets), our analysis is compatible with other recent analysis
on $w_Q$ (\cite{others}).
Our final result is perfectly in agreement with the $w_Q=-1$ 
cosmological constant case and gives no support to a 
quintessential field scenario with $w_Q > -1$.
A frustrated network of domain walls or
a purely exponential scaling field are excluded at high 
significance. In addition a number of quintessential models 
are highly disfavored, power law potentials with $p\ge 1$ and the oscillatory 
potential discussed in this paper, to name a few.

It will be the duty of higher redshift datasets, for example
from clustering observations \cite{calvao} to point to 
a variation in $w$ that might place quintessence in a more 
favorable light. 

The result obtained here, however, could be plagued by some
of the theoretical assumptions we made.
The CMB and LSS constraints can be weakened by the inclusion of a 
background of gravity waves or of isocurvature perturbations or 
by adding features in the primordial perturbation spectra.
These modifications are not expected in the most basic 
and simplified inflationary scenario but they are still 
compatible with the present data.
The SN-Ia result has been obtained under the assumption of a 
constant-with-time $w_Q$. 
We have shown that in general $w_{eff}$ is
a rather good approximation for dynamical quintessential mode\
ls since the luminosity distance depends on
$w_Q$ through a multiple integral that smears its redshift 
dependence. As we show in the previous section, our result is
therefore valid for a wide class of quintessential models.
This `numbing' of sensitivity to $w_Q$ implies that maybe 
an effective equation of state is the most tangible parameter 
able to be extracted from supernovae. However with the promise of 
large data sets from Planck and SNAP satellites, 
opportunities may yet still be open to reconstruct a time varying 
equation of state \cite{jochen}.

\medskip

\textit{Acknowledgements} It is a pleasure to thank 
Ruth Durrer, Steen Hansen and Matts Roos
for comments and suggestions. RB and AM are supported by PPARC.
We acknowledge the use of CMBFAST~\cite{CMBFAST}.

\end{document}